\documentclass[12pt,a4paper,final]{iopart}
\usepackage{graphicx}

\usepackage{iopams}  
\usepackage[breaklinks=true,colorlinks=true,linkcolor=blue,urlcolor=blue,citecolor=blue]{hyperref}
\usepackage[utf8]{inputenc}
\usepackage[english]{babel}
\expandafter\let\csname equation*\endcsname\relax
\expandafter\let\csname endequation*\endcsname\relax
\usepackage{amsmath}

\begin{document}
\title[]{Disalignment rate coefficient of argon $\rm 2p_8$ due to nitrogen collision}
\author{Roman Bergert, Leonard W. Isberner, Slobodan Mitic, Markus H. Thoma}
\address{Institute of Experimental Physics I, Justus-Liebig-University Giessen Heinrich-Buff-Ring 16,
D-35392 Giessen, Germany}
\ead{Roman.Bergert@physik.uni-giessen.de}

\begin{abstract}

Tunable diode laser induced fluorescence (TDLIF) measurements are discussed and quantitatively evaluated for nitrogen admixtures in argon plasma under the influence of a strong magnetic field. TDLIF measurements were used to evaluate light-transport properties in a strongly magnetized optically thick argon/nitrogen plasma under different pressure conditions. Therefore, a coupled system of rate balance equations was constructed to describe laser pumping of individual magnetic sub-levels of $\rm 2p_8$ state through frequency-separated sub-transitions originating from $\rm 1s_4$ magnetic sub-levels. The density distribution (alignment) of $\rm 2p_8$ multiplet was described by balancing laser pumping with losses including radiative decay, transfer of excitation between the neighboring sub-levels in the $\rm 2p_8$ multiplet driven by neutral collisions (argon and nitrogen) and quenching due to electron and neutral collisions. Resulting $\rm 2p_8$ magnetic sub-level densities were then used to model polarization dependent fluorescence, considering self-absorption, which could be directly
compared with polarization-resolved TDLIF measurements. This enables to estimate the disalignment rate constant for the $\rm 2p_8$ state due to collisions by molecular nitrogen. A comparison to molecular theory description is given providing satisfactory agreement.
The presented measurement method and model can help to describe optical emission of argon and argon-nitrogen admixtures in magnetized conditions and provides a basis for further description of optical emission spectra in magnetized plasmas.
\end{abstract}

\noindent{\it Keywords\/}: {magnetized plasma diagnostics, plasma jet, dielectric barrier discharge (DBD), low-pressure plasma, magnetized plasma, magnetic sub-level population, argon nitrogen plasma, laser-induced fluorescence (LIF), disalignment rate constant}

\maketitle
%\twocolumn
\section{Introduction}

Quantitative evaluation of plasma emission spectra under the influence of a magnetic field is still limited by the theoretical description of interactions within multiplets of energetic sub-levels created by the Zeemann effect \cite{zeeman1897xxxii}. Splitting of energetic levels will lead to an additional set of inter-multiplet interactions (excitation, radiation) between the sub-levels of different multiplets that will define the escaped plasma radiation \cite{Fujimoto1988, Matsukuma2012}. Besides the excitation and radiation interactions that should be adjusted between the multiplets, additional interaction has to be considered describing the non-radiative transition of excitation within the multiplet of a single level (intra-multiplet transitions) or between multiplets of different 2p or 1s energetic states (inter-multiplet transitions). This way, the rate balance equations describing the population density within an excited multiplet will have to include the transfer of excitation from other sub-levels driven by collisions with neutrals.  

Inter- and intra-multiplet transitions in argon plasma are rarely discussed in literature. Most previous literature provide information regarding pure argon and admixture with other noble gases (helium, neon, krypton), at different temperatures and several 2p states (in Paschen notation) \cite{Matsukuma2012,grandin1973sections,grandin1978depolarisation,Grandin1981,bergert2020quantitative}. Besides the quite limited number of articles evaluating disalignment properties of argon with other noble gas perturbers, the effect of other atomic and molecular species is not discussed in literature yet.

Intra-multiplet transitions strongly depend on the perturber which defines the population distribution (alignment) of the multiplet \cite{Matsukuma2012}. Intra-multiplet transition between adjacent levels, responsible for equalizing the population within the multiplet and thus destruction of an alignment, is described by the disalignment constant.
The disalignment rate will have a direct impact on the optical properties of a plasma by influencing the sub-state density distribution within the Zeeman-splitted multiplet and thus polarization of plasma emission lines which will be reflected in the total line intensity escaping the plasma.

In this work, a tunable diode laser was used to individually target sub-transitions between $\rm 1s_4$ and $\rm 2p_8$ multiplets of argon separated by a frequency shift due to the magnetic field. Laser absorption measurements were used to evaluate the $\rm 1s_4$ and $\rm 1s_5$ sub-level densities. These results were later used in the description of laser induced fluorescence measurements. 
The disalignment rate coefficient of $\rm 2p_8$ state was evaluated in pure argon plasma in our previous work \cite{bergert2020quantitative}. The evaluated disalignment rate coefficient for argon is further used for the description of $ \rm Ar/N_2$ mixtures in order to extract the disalignment rate driven by collisions with nitrogen molecules. The estimation of a disalignment rate coefficient is strongly dependent on the precision of the estimated number density of the perturbers in the plasma cell where direct measurements were not feasible. This problem was solved by measuring Doppler shift induced by fast gas flow velocity through the cell. Doppler shift was measured by laser induced fluorescence providing the estimation of total pressure in the cell that could be correlated with the partial pressure of argon and nitrogen perturbers. Finally the estimated disalignment constant for nitrogen molecules is compared with theoretical estimations providing satisfactory agreement.

\section{Methods}
\label{methods}

\begin{figure}[h]
	\centering
	\includegraphics[width=0.5\linewidth]{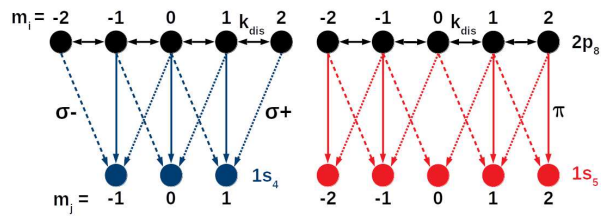}
	\caption{Kastler diagram indicating possible optical transitions from upper level $\rm 2p_8$ to lower $\rm 1s_4$ and $\rm 1s_5$ levels under the influence of an external magnetic field, as well as redistribution of $\rm 2p_8$ sub-levels due to disalignment.}
	\label{fig:kastler}
\end{figure}

Under the influence of a magnetic field, each energetic level will split into a multiplet with 2J+1 sub-levels which are described by the magnetic quantum numbers $m$ ranging from -J to J. A higher energetic multiplet (e.g. $\rm 2p_8$) produces a system of sub-transitions to a lower energetic multiplet (e.g. $\rm 1s_4$) defined by the optical transition rules. In such configuration, changes of the magnetic quantum number $\rm \Delta m=0$ are creating linear ($\pi$) polarized while $\rm \Delta m=\pm 1$ will result in circular ($\sigma\pm$) polarized emission. Sub-transitions are shifted in frequency by the external magnetic field producing a symmetric absorption structure around the original unshifted line center without an external magnetic field. The allowed electric dipole transitions originating from $\rm 2p_8$ to $\rm 1s_4$ and $\rm 1s_5$ are shown in figure \ref{fig:kastler} by their polarization. The $\rm 2p_8$ (J=2) and $\rm 1s_5$ (J=2) states split into five sub-levels. Hence, twelve different electric dipole transitions are formed (4 $\sigma+$, 4 $\pi$ and 4 $\sigma-$), which is indicated with red arrows between the sub-levels of $\rm 2p_8$ and $\rm 1s_5$. Nine sub-transitions (3 $\sigma+$, 3 $\pi$ and 3 $\sigma-$), which are indicated with blue arrows in figure \ref{fig:kastler}, can be found between $\rm 2p_8$ and $\rm 1s_4$ (J=1). 
Redistribution of upper $\rm 2p_8$ sub-levels caused by disalignment is shown by black arrows in figure \ref{fig:kastler}.\\
In this work, tunable diode laser spectroscopy was used in various configurations providing different information about the system. Due to a relatively strong external magnetic field of 0.3 T, targeted sub-transitions between $\rm 1s_4$ and $\rm 2p_8$ sub-levels were well-separated by frequency so that the laser could be tuned to pump each transition individually. According to the set of Einstein coefficients describing sub-transition strength derived in \cite{bergert2020quantitative}, the strength of the laser induced interaction can be quantitatively described. 

In absorption mode, the laser scan was set for a wide frequency range scanning through all possible sub-levels of the measured 1s multiplet as described in \cite{Bergert_2019}. As a result, the absolute state density of each targeted 1s multiplet could be reconstructed providing additionally the alignment of the multiplet. LAS measurements were done on $\rm 1s_4$ and $\rm 1s_5$ multiplets for a set of measurement conditions with different pressures and nitrogen concentrations.

The description of tunable diode laser induced fluorescence (TDLIF) measurements was introduced for the pure argon case in our previous article \cite{bergert2020quantitative} and it will be briefly described in the following section. Theoretical calculation of the disalignment cross section will be introduced which will be further used to calculate the theoretical disalignment rate coefficient for Ar/$\rm N_2$, and compared with experimentally obtained values. Section \ref{methods} will end with the description of the pressure estimation in the plasma cell, based on the Doppler shift induced by the gas flow, measured by TDLIF.

\subsection{Tunable diode laser induced fluorescence in magnetized plasma}

The laser induced fluorescence measurements were based on the excitation of different sub-transitions between $\rm 1s_4$ and $\rm 2p_8$ multiplets by a wide frequency scan ($\rm \sim 8 GHz$) around the central transition at 842.47 nm. Fluorescence was observed by a spectrometer at 842.47 nm and 801.48 nm. The linear and circular polarized parts of the fluorescence of both lines could be measured through a polarization filter isolating $\rm \pi$ or $\rm \sigma$ ($\sigma+$ and $\sigma-$) polarization depending on the orientation of the filter.  

The efficiency of laser pumping for each transition is therefore proportional to the laser intensity, the density of the targeted $\rm 1s_{4,m_j}$ state and the Einstein coefficient for absorption of the pumped sub-transition. In such configuration, considering polarization of the laser light, each transition interacts only with the corresponding polarity from the unpolarized pumping laser so that laser intensities for different polarizations should scale as $I_\pi=2I_\sigma$.

Pumping of an individual $\rm 2p_8$ sub-level would lead to polarization dependent fluorescence emission $F^\lambda_\pi$ and $F^\lambda_\sigma$, determined by the ratio of Einstein coefficients of the $\pi$ and $\sigma$ components originating from the pumped sub-level (and other sub-levels populated by disalignment) and corrected for self-absorption of $\rm \lambda=842.47$ and $\rm 801.48 \, nm$ sub-transitions, respectively. This way pumping of each of the $\rm 2p_8$ sub-levels through different sub-transitions would result in a unique polarization of induced fluorescence on both wavelengths building the system of non linear equations describing the TDLIF. 

Laser induced fluorescence was modeled including the contribution of collisional and radiative processes to the final state density distribution of the excited $\rm 2p_8$ sub-levels. 
As proposed in \cite{bergert2020quantitative}, polarized fluorescence intensities at both measured branches, $\rm \lambda=842.47, 801.48 \, nm$, can be modeled by taking into account Einstein coefficients for spontaneous emission $A^\lambda_{m_i,m_j}$ for each branch and each excited sub-transition, with upper level magnetic quantum number $m_i$ and lower level magnetic quantum number $m_j$, as well as self-absorption coefficients $\gamma^\lambda_{m_i,m_j}(n^{\rm 1s_2,1s_4,1s_5}_{m_j})$ depending on $\rm 1s_2$, $\rm 1s_4$ or $\rm 1s_5$ sub-level densities $n^{\rm 1s_2,1s_4,1s_5}_{m_j}$.
In addition to the optical interactions (radiation, self-absorption and laser pumping) there is the collisional part which, through the disalignment process, can significantly influence the resulting density distribution within $\rm 2p_8$ multiplet in the TDLIF measurements. Disalignment is included in the rate balance equation of each $\rm 2p_8$ sub-level as a production and loss source due to the intra-multiplet transitions. This way, the disalignment process is competing with laser pumping in production of an excited $\rm 2p_8$ sub-level population while at the same time competing with radiative decay in destruction of the excited sub-level density. According to \cite{Matsukuma2012,Matsukuma2011,Seo_2003} the intra-multiplet transition can only happen between the neighboring states with the same probability.

The resulting densities can be described by steady state rate balance equations, depending on whether laser pumping to such a level takes place.
If a level Ar($\rm 2p_{8,m_i}$) is not directly pumped by laser, the rate balance equation describing the density in steady state solution is given by
\begin{align}
n^{2p_8}_{m_i} = \frac{c_k \cdot \sum_{\lvert m_i-m_j \rvert = 1} n^{2p_8}_{m_j}}{a \cdot c_k + \tau_{m_i} + c_q}.
\label{eq:dens}
\end{align}
In this case, the sub-state $\rm 2p_{8,m_i}$ is assumed to be populated only by disalignment from adjacent $\rm 2p_8$ sub-states with $\Delta m =\pm 1$.
This procedure is expressed by the product of neighboring state densities $n^{\rm 2p_8}_{m_j}$ and disalignment rate $c_k$, which is expected to be the same for all transitions with $\Delta m =\pm 1$.

For an argon/nitrogen plasma, the disalignment rate can be separated into rates for both gas species Ar and $\rm N_2$,
\begin{align}
	c_k = n_\text{Ar} \cdot k_\text{dis,Ar} + n_{\text{N}_2} \cdot k_\text{dis,N$_2$},
	\label{eq:ck}
\end{align}
that are given as products of neutral densities $n$ of the gas components respectively and gas-specific disalignment constants $k_\text{dis}$.
The disalignment constant for argon $\rm 2p_8$ state $k_\text{dis,Ar}$, reported in \cite{grandin1973sections} as $1.3\times 10^{-15}\, \mathrm{\frac{m^3}{s}}$ was also confirmed within previous work \cite{bergert2020quantitative}.
For the determination of $k_\text{dis,N$_2$}$ in this paper, $k_\text{\rm dis,Ar} = 1.3\times 10^{-15}\, \mathrm{\frac{m^3}{s}}$ was used for all experimental conditions.
Argon and nitrogen neutral densities ($n_{\rm Ar}$ and $n_{\rm N_2}$) were calculated from pressure and gas flow measurements further described in the following subsection \ref{subsection:pressure}.

Depopulation of $\rm 2p_8$ multiplet happens via three processes, taking into account intra-multiplet transitions to other $\rm 2p_8$ sub-levels, radiative decay to 1s states and excitation transfer to other 2p and 1s multiplets caused by collisions with neutrals and electrons (inter-multiplet transitions).

Transfer within the magnetic $\rm 2p_8$ multiplet is determined by the disalignment rate and the number $a$ of neighboring $\rm 2p_8$ states where $a=1$ for $m_i=\pm 2$ and $a=2$ for $m_i = 0, \pm 1$.
The inverse radiative lifetime $\tau_{m_i}$ describes all possible decay branches of $\rm 2p_{8, m_i}$ levels, comprising transitions to $\rm 1s_4$, $\rm 1s_5$, and $\rm 1s_2$ ($\lambda =$ 978.45 nm) multiplets, and is given by

\begin{align}
\tau_{m_i} &= \tau^{842}_{m_i} + \tau^{801}_{m_i} + \tau^{978}_{m_i} \\
&= \sum_{\substack{m_j=0,\pm 1\\ \lvert m_i-m_j \rvert \leq 1}} A^{842}_{m_im_j} \cdot \gamma^{842}_{m_im_j}
		+ \sum_{\substack{m_j=0,\pm 1, \pm 2\\ \lvert m_i-m_j \rvert \leq 1\\ \lvert m_i+m_j \rvert \geq 1}} A^{801}_{m_im_j} \cdot \gamma^{801}_{m_im_j}
		+ \sum_{\substack{m_j=0,\pm 1\\ \lvert m_i-m_j \rvert \leq 1}} A^{978}_{m_im_j} \cdot \gamma^{978}_{m_im_j}
\end{align}

The third loss mechanism describes quenching with neutrals and electrons leading to excitation transfer to other 2p and 1s multiplets in plasma.
To consider this process, a global quenching rate $c_q$ is introduced,
\begin{align}
	c_q = n_\text{Ar} \cdot k_\text{q,Ar} + n_{\text{N}_2} \cdot k_\text{q,N$_2$} + n_e \cdot k_{q,e},
\end{align}
representing collisions with argon neutrals, nitrogen molecules and electrons.
The quenching rate is given as product of species density $n$ and deactivation rate constant $k_\text{q}$, reported in \cite{Chang1978} for $k_\text{q,Ar}$, \cite{sadeghi2001} for $k_{\rm q,N_2}$ and \cite{zhu2010} for $k_{q,e}$ .
The electron density for argon admixture with nitrogen was expected to be in order of few $10^{16}\, \rm m^{-3}$ comparable to measured values in \cite{kaupe2019phase} for unmagnetized plasma and similar conditions.

If the laser frequency matches the transition frequency $\rm 1s_{4,m_j} \rightarrow 2p_{8,m_i}$, the sub-level is pumped and its density can be obtained by
\begin{align}
n^{2p_8}_{m_i} = \frac{\left(c_k \cdot \sum_{\lvert m_i-m_j \rvert = 1} n^{2p_8}_{m_j}\right) + n^{1s_4}_{m_j} \cdot I_{\pi / \sigma} \cdot B^{842}_{m_i, m_j}}{a \cdot c_k + c_q + \tau_{m_i} + \gamma^{842}_{m_i, m_j} \cdot I_{\pi / \sigma} \cdot B^{842}_{m_j, m_i}}.
\label{eq:lasdens}
\end{align}
The described loss and gain mechanisms (see (\ref{eq:dens})) are still valid, while additional terms taking into account interaction of laser light with particles will have to be added.
Excitation of the $\rm 2p_{8,m_i}$ sub-state by laser pumping from the lower $\rm 1s_{4,m_j}$ sub-state can be expressed as a product of $\rm 1s_4$ state density $n^{\rm 1s_4}_{m_j}$, laser light intensity with corresponding polarization $\rm I_{\pi,\sigma}$ and Einstein coefficient for absorption $\rm B^{842}_{m_i,m_j}$.
As the laser light is unpolarized, not every photon can induce the desired excitation where only linear ($\rm I_\pi$) or circular ($\rm I_{\sigma}$) polarized component can induce $\Delta m=0$ or $\Delta m=\pm 1$ transitions respectively.

An additional term in (\ref{eq:lasdens}) is introduced as a loss mechanism describing the laser induced emission from pumped $\rm 2p_8$ sub-state, proportional to the laser intensity and the Einstein coefficient for induced emission $\rm B^{842}_{m_j,m_i}$. Due to the high laser intensity in the order of $\rm kW/m^2$, this additional loss channel was found to be around 10\% compared to all other loss processes.

For each laser pumped sub-transition, (\ref{eq:dens}) and (\ref{eq:lasdens}) can be used to describe a steady state density distribution within $\rm 2p_8$ multiplet which can be further used to compare modeled and measured polarization dependent fluorescence for 842.47 and 801.48 nm transitions. The intensity of each polarity originating from an excited $\rm 2p_8$ multiplet can be expressed as a sum of the sub-transitions with corresponding polarities, each proportional to the originating $\rm 2p_8$ sub-state density and Einstein coefficient for spontaneous emission $A^{\lambda}_{m_i,m_j}$, corrected for self-absorption. Detailed description of each polarization component for both 842.47 and 801.48 nm fluorescence branches can be found in \cite{bergert2020quantitative}.

\subsection{Theoretical estimation of the disalignment rate coefficient}
\label{disrate}

According to previous works \cite{Matsukuma2012,Carrington_1971}, the disalignment effect can be theoretically expressed in form of cross-section of alignment destruction as

\begin{equation}
    \sigma^{(x)}_J=\phi^{(x)}_J\frac{<v^{3/5}>}{<v>}\left(\frac{\pi p^2 p^2_i}{16<\Delta E>}\right)^{2/5},
    \label{eq:sigma}
\end{equation}
where $x$ is the $2^x$-pole moment which is in our case $x=2$ for alignment. The factor $\phi^{(x)}_J$ is a geometrical factor depending on pole and total angular momentum quantum number and can be found in \cite{Wang1969}. The eigenvalue of the electric dipole momentum of the perturbing atom or molecule is $p^2_i$ and can be expressed as the static dipole polarizability $|\alpha|$, which can be found for $\rm N_2$ in \cite{Spelsberg1994}. $p$ is the electric dipole momentum of the perturbed atom, in our case 18.08 in atomic units for argon \cite{grandin1973sections}. The average energy $<\Delta E>$ represents the difference of the perturbing-perturbed atom system before collision and can be approximated as the first excited level of the perturbed atom \cite{Matsukuma2012}. In our case $<\Delta E>=11.83$ eV which is the energy of argon $\rm 1s_2$ state. According to \cite{Carrington_1971}, the expression
\begin{equation}
    \frac{<v^{3/5}>}{<v>}=0.9314\left(\frac{\mu}{2k_BT}\right)^{\frac{1}{5}}
\end{equation} 
depends on the reduced mass $\mu=\frac{M_1 \, M_2}{M_1+M_2}$, the neutral gas temperature T and the Boltzmann constant $k_B$. Finally, the disalignment rate constant can be calculated as $k_{\rm dis,N_2}=<v>\sigma^{(2)}_J$ with $<v>=\sqrt{\frac{8k_BT}{\pi\mu}}$ as average velocity of the ensemble.

\subsection{Pressure estimation}
\label{subsection:pressure}

The total pressure $p_\text{tot}$ inside the plasma jet was estimated from the Doppler shift induced by the gas flow velocity measured by TDLIF. Measurements were performed in the unmagnetized case targeting central $\rm 1s_4 \rightarrow \rm 2p_8$ transition with the laser directed along the gas flow in axial direction of the discharge tube. The maximum of TDLIF and maximum peak position of the radial absorption profile were used to measure the induced Doppler shift as explained in more detail in \cite{kaupe2018phase,mitic2019comparative}.
For an argon/nitrogen plasma, total pressure consists of partial pressures for argon and nitrogen,
\begin{align}
	p_\text{tot} = p_\text{Ar} + p_\text{N$_2$}.
\end{align}
Assuming the same gas speed and temperature for argon and nitrogen in the tube, partial pressures for a gas type (Ar or $\rm N_2$) can be estimated from gas flows $f$ as
\begin{align}
	p_\text{gas} &= \frac{f_\text{gas}}{f_\text{Ar} + f_\text{N$_2$}} \cdot p_\text{tot}
\end{align}

According to the ideal gas law, the neutral densities of each component were obtained by
\begin{align}
n_\text{gas} = \frac{N_\text{gas}}{V} = \frac{p_\text{gas}}{k_\text{B} \, T_\text{gas}}.
\label{eq:neutraldens}
\end{align}
Temperature in the plasma tube was 340 K which is estimated based on LAS measurements presented in previous work \cite{mitic2019comparative}.

\section{Experimental Setup}

\begin{figure}[h]
	\centering
	\includegraphics[width=0.5\linewidth]{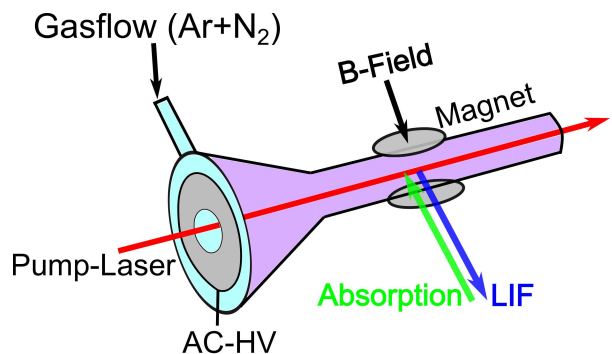}
	\caption{Schematic representation of the experimental setup.}
	\label{fig:set-up}
\end{figure}

The experimental setup and discharge configuration is presented in figure \ref{fig:set-up}. It consists of a cone shaped glass discharge chamber followed by a tube of 4 mm inner diameter creating a compact plasma jet. The open end of the tube was attached to a 500 mm long expansion chamber with the vacuum system attached at the end. The driven electrode made of aluminum tape was attached from the outside of the cone base with an opening in the center to introduce a laser beam in axial direction. Laser paths are indicated with a red arrow for TDLIF and a green arrow for absorption measurements. Fluorescence induced by the axial laser was detected perpendicularly as indicated with the blue arrow.

A base pressure of 1.4$ \times 10^{-1}$ Pa measured inside the expansion chamber could be reached.
Plasma was created by a bipolar 30 kHz sinusoidal high voltage ($V_{pp}\approx$ 4 kV) signal. Two cylindrical permanent magnets of 6 mm in diameter were mounted on opposite sides above and below the discharge tube as shown in figure \ref{fig:set-up}.
The induced external magnetic field was perpendicular to the line of sight and had a strength  of 0.3 T as measured by a Gaussmeter (\textit{LakeShore}).

Laser absorption measurements were done in radial direction crossing the tube between the magnets, including common optical elements like an argon reference cell and a Fabry-P\'{e}rot interferometer to monitor laser scanning range and quality. Absorption at 842 nm transition was scanned by a \textit{TOPTICA DLC 100} laser and at 801 nm by \textit{TOPTICA DLC pro}. A system of a collimators and a multimode $200~\mathrm{\mu m}$ optical fiber was used to manipulate the laser beams, producing an unpolarized probing laser beam. Two band pass filters at 800 nm and 840 nm with a full width at half-maximum of 10 nm were used in order to suppress the unwanted plasma emission. With the absorption measurements organized in such configuration (orientated perpendicular to the magnetic field lines) it is possible to probe transitions with both polarizations.

For fluorescence measurements, laser light was introduced perpendicular to the external magnetic field in axial direction of the discharge tube along the gas flow direction. The three $\pi$ and three $\sigma\pm$ transitions induced from $1s_4$ at 842 nm were used to consecutively pump the $\rm 2p_8$ sub-levels. Laser scan was set at low scanning frequency ($\rm \sim 0.1\, Hz$) allowing simultaneous measurements of induced fluorescence with a high sampling rate (20-85 ms integration time) using a $200~\mathrm{\mu m}$ optical fiber and a collimator connected to an Ocean Optics USB2000+ spectrometer. The fluorescence was observed perpendicular to the illumination laser using identical optics as for the LAS measurements.
A linear polarization filter from Thorlabs GmbH with a total transmission of 40 \% for unpolarized light was used between plasma and collection optics to isolate desired polarization from the induced fluorescence. With the linear polarization filter oriented parallel to the magnetic field lines, only $\pi$ transitions could be transmitted, while perpendicular orientation would transmit only $\sigma$. The polarization filter was not completely isolating the desired polarity so that 3 \% of transmissions of the opposite polarity was measured and was further taken into account for correct reconstruction of polarization dependent intensity. Relative sensitivity for linear and circular polarized light was evaluated to be 1:0.7, which is usually dependent on the optics and the detector system.

\section{Results and Discussion}
\label{sec:resultsanddiscussion}

The experimental measurements were done using two different argon gas flow settings of 55.6 and 83.4 sccm, and several admixtures of nitrogen (5, 15, 30, 50 sccm) providing different experimental conditions for the estimation of the Ar($\rm 2p_8$) disalignment rate coefficient due to nitrogen collisions.
Laser spectroscopic measurements explained in section \ref{methods} were done for each experimental condition.
In the first step, gas pressure was evaluated by axial TDLIF measurements similar to \cite{kaupe2018phase,mitic2019comparative}, targeting argon $\rm 1s_4$ state, in unmagnetized conditions. Based on the difference of the line centers measured by TDLIF and radial TDLAS, gas velocity between the magnets could be correlated to the gas pressure due to the continuity of the gas flow.
\begin{figure}[h]
	\centering
	\includegraphics[width=.5\linewidth]{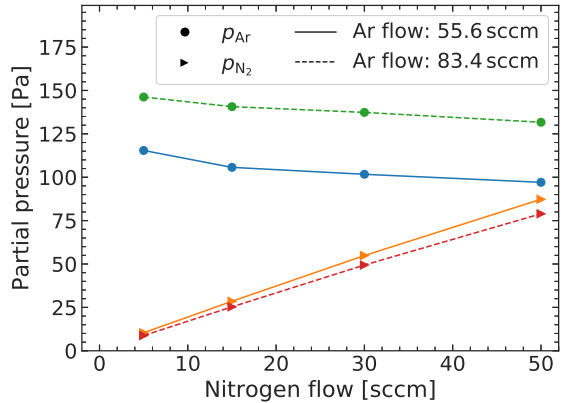}
	\caption{Argon and nitrogen partial pressures at different gas flows.}
	\label{fig:pressure}
\end{figure}
By this way, the total pressure inside the tube could be evaluated and further used to estimate the partial pressures of each gas component for all mixtures, as presented in figure \ref{fig:pressure}.

Further, radial TDLAS measurements were done in magnetized conditions, targeting $\rm 1s_4$ and $\rm 1s_5$ multiplets, providing information about the 1s sub-state absolute densities for each gas mixture. The absorption measurements under similar conditions were already described in our previous work \cite{Bergert_2019}, where detailed description of the measurement procedure and evaluation were explained. The evaluated $\rm 1s_4$ and $\rm 1s_5$ sub-state densities for each condition are further used in the modeling of measured fluorescence in order to reduce the number of unknowns and increase the accuracy of the estimated disalignment effect driven by nitrogen collisions. 

\begin{figure}
	\centering
	\includegraphics[width=.5\linewidth]{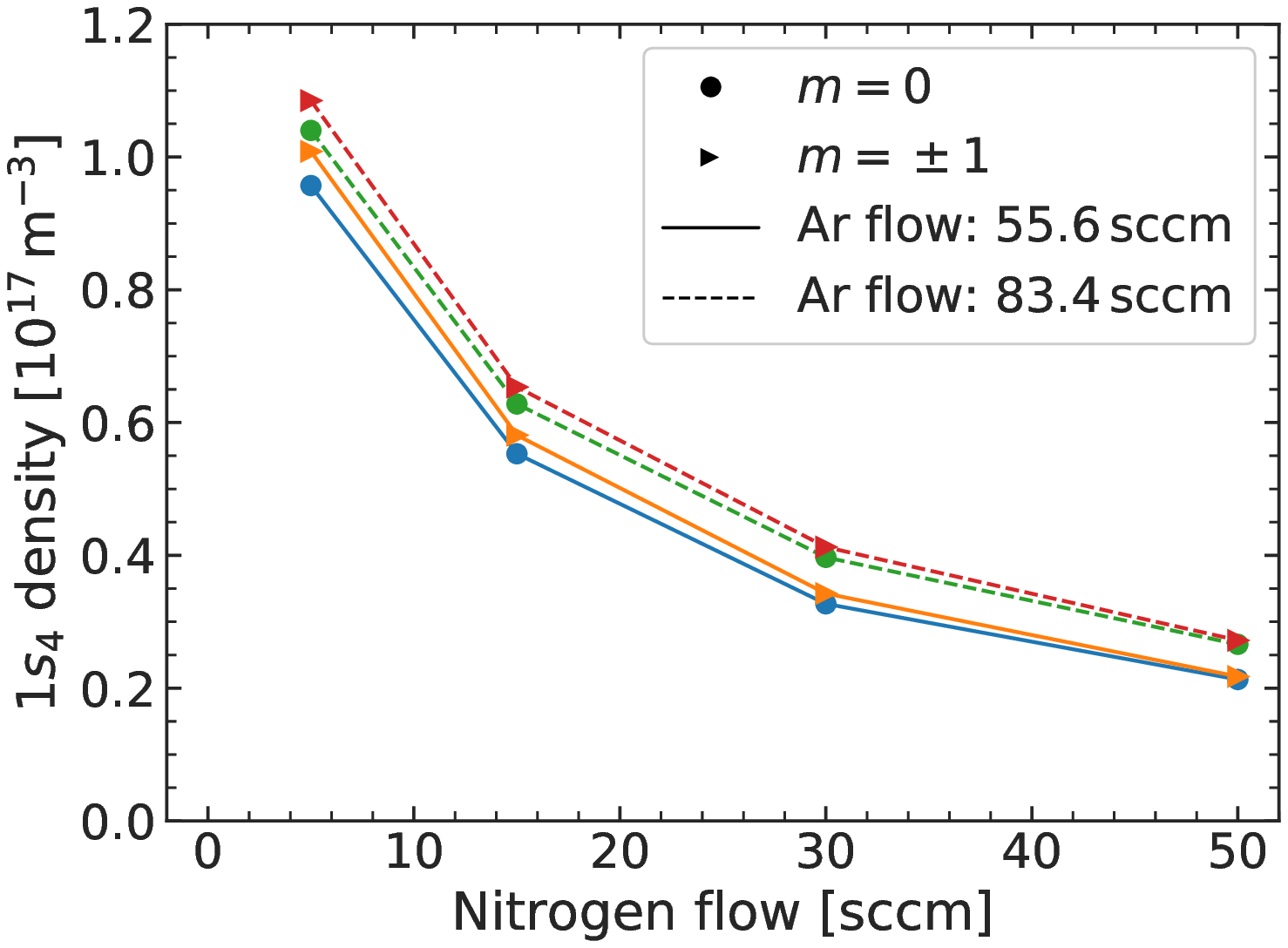}
	\caption{Measured $\rm 1s_4$ magnetic sub-level densities ($m=0,\pm1$) for different argon and nitrogen flows.}
	\label{fig:dens1s4}
\end{figure}

\begin{figure}
	\centering
	\includegraphics[width=.5\linewidth]{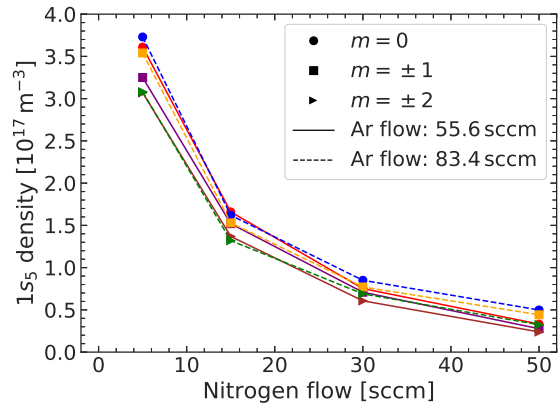}
	\caption{Measured $\rm 1s_5$ magnetic sub-level densities ($m=0,\pm 1, \pm 2$) for different argon and nitrogen flows.}
	\label{fig:dens1s5}
\end{figure}

The evaluated $\rm 1s_4$ and $\rm 1s_5$ multiplet densities for all together eight different gas mixtures are presented in figures \ref{fig:dens1s4} and \ref{fig:dens1s5} respectively.
As already found in \cite{Bergert_2019}, both $\rm 1s_4$ and $\rm 1s_5$ multiplets have symmetric positive and negative alignment respectively. This fact does allow to reduce the number of variables in modeling of fluorescence by using identical densities for sub-states with opposite signs in magnetic quantum numbers ($n_{m_j}=n_{-m_{j}}$).

A significant drop of an order of magnitude in 1s state densities was observed when nitrogen concentration was increased up to 50 sccm. Similar behavior was observed in previous work \cite{kaupe2019phase} based purely on optical emission measurements and collisional-radiative modeling.
Precise modeling of $\rm 2p_8$ state requires the description of one more transition branch $\rm 2p_8 \rightarrow 1s_2$ at 978.45 nm. However, due to a low Einstein coefficient for this transition and a low $\rm 1s_2$ state density \cite{kaupe2018phase}, the inclusion of the 978.45 nm branch does not strongly contribute to the population density of $\rm 2p_8$ state compared to other branches. According to \cite{kaupe2018phase}, $\rm 1s_2$ sub-densities were assumed as roughly one third of $\rm 1s_4$ densities.

\begin{figure}[h]
	\centering
	\includegraphics[width=.5\linewidth]{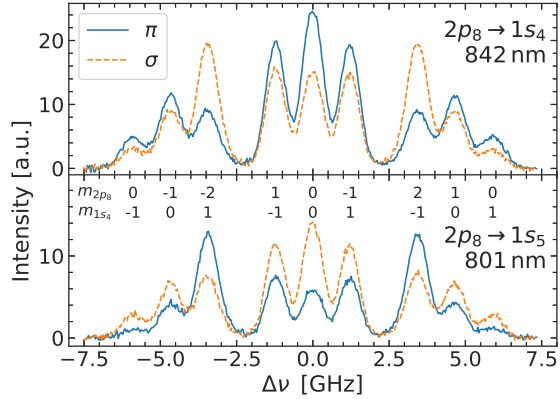}
	\caption{LIF profiles for 55.6 sccm of argon and 5 sccm of nitrogen at the transition wavelength for $\rm 2p_8$ to $\rm 1s_4$ (top) and $\rm 1s_5$ (bottom), respectively. $\Delta \nu$ denotes the detuning of the pumping laser compared to the unshifted transition frequency for $\rm 1s_4, m=0 \rightarrow 2p_8, m=0$. $\pi$-polarized fluorescence is marked as solid blue line, $\sigma$-polarized light as dashed orange line. Small numbers between the profiles are indicating the pumped transition.}
	\label{fig:lif_signal}
\end{figure}

In case of TDLIF measurements, laser frequency was set to scan over a wide frequency range so that all nine possible transitions from $\rm 1s_4$ to $\rm 2p_8$ magnetic sub-levels could be pumped successively.
Linear and circular polarized fluorescence light was recorded one after the other by rotating the polarization filter by $90^{\circ}$.
Emission from $\rm 2p_8 \rightarrow 1s_4$ at 842.47 nm and from $\rm 2p_8 \rightarrow 1s_5$ at 801.48 nm was measured simultaneously using a spectrometer. The resolution of the spectrometer was not high enough to separate magnetically shifted wavelength, so the individual sub-transitions originating from $\rm 2p_8$ multiplet could not be resolved. Hence, the observed fluorescence for a fixed polarization and pumping channel is always a sum of all possible sub-transitions origination from all $\rm 2p_8$ sub-levels. Due to the disalignment effect, the population of the pumped $\rm 2p_8$ sub-level and all others are distributed so that each possible sub-transition has to be taken into account for all pumped sub-levels.
Recorded data was then corrected for plasma background.
An example of a resulting laser induced fluorescence profile is presented in figure \ref{fig:lif_signal} separated for each wavelength.
A nine-fold structure produced by separately pumping magnetic sub-levels is clearly visible and transitions can be identified by the expected shift in frequency compared to the unshifted central line corresponding to $\rm 1s_4, m=0 \rightarrow 2p_8, m=0$ sub-transition.
The three peaks in the center are created by inducing $\pi$ transitions while the three peaks on the left and right are produced by inducing $\sigma +$ and $\sigma -$ transitions respectively.

For each pumped transition, (\ref{eq:dens}) and (\ref{eq:lasdens}) were used to describe the population density of the $\rm 2p_8$ states, depending on whether the state is directly pumped or indirectly populated purely by the disalignment process. The resulting set of equations could be solved providing information about density distribution within the $\rm 2p_8$ multiplet caused by laser pumping of a certain transition. The multiplet density distribution was further used to model the escaped radiation taking into account $\rm 1s_4$ and $\rm 1s_5$ multiplet densities previously obtained by TDLAS. As a result for each excited transition, the polarization of induced fluorescence at 842 and 801 nm could be analytically described.

To find the desired disalignment rate coeffient $k_{\rm dis,N2}$, the total disalignment rate $c_k$ was determined with a least-squares fitting routine for all measured mixtures of argon and nitrogen separately.
Therefore, the expected polarization-dependent fluorescence intensities $M$ for all pumping sub-transitions $1s_4 \rightarrow 2p_8$ were modeled including known parameters as 1s sub-level densities, partial pressures, neutral gas temperature and laser intensity. The ratios of circular and linear polarized modeled fluorescence $\frac{M_\sigma}{M_\pi}(c_k, m_i, m_j)$  when pumping a particular transition $m_j \rightarrow m_i$ were adapted by varying $c_k$ to fit the ratios of corresponding polarization-dependent intensity peaks $F$ in measured TDLIF profiles best.
Thus, the differences of ratios of maximum intensities in TDLIF measurements and modeled fluorescence generate a set of nine equations for each fluorescence branch,
\begin{align}
	\frac{M_\sigma}{M_\pi}(c_k, m_i, m_j) - \frac{F_\sigma}{F_\pi}(m_i, m_j) \stackrel{!}{=} 0,
\end{align}
 which were minimized collectively to obtain a suitable $c_k$. 

\begin{figure}
	\centering
	\includegraphics[width=.5\linewidth]{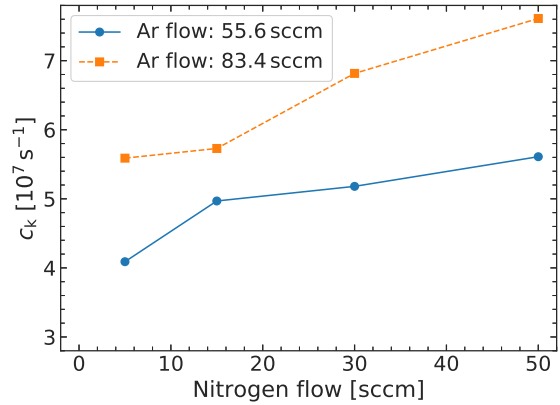}
	\caption{Values for disalignment rate $c_k$ as obtained in the fitting process.}
	\label{fig:ck}
\end{figure}

Obtained values for disalignment rate coefficient are plotted in figure \ref{fig:ck}.
Fitting errors were in the range between 5--19 $\%$.
As expected, $c_k$ increases when more nitrogen is injected and is always higher for a higher argon flow rate.

\begin{figure}
	\centering
	\includegraphics[width=.5\linewidth]{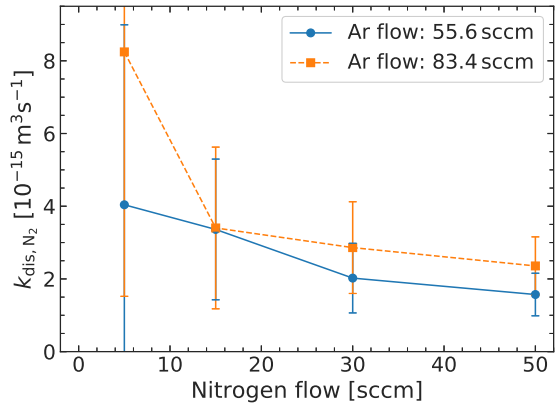}
	\caption{Disalignment constant $k_\text{dis,N$_2$}$ as function of nitrogen flow.}
	\label{fig:kdis}
\end{figure}

Finally, the calculated disalignment rates $c_k$ for admixtures of nitrogen were used to extract nitrogen disalignment rate coefficient $k_\text{dis,N$_2$}$, given by
\begin{align}
	k_\text{dis,N$_2$} = \frac{k_\text{B} \, T \cdot c_k - p_\text{Ar} \cdot k_\text{dis,Ar}}{p_\text{N$_2$}}.
\end{align}
Results are shown in figure \ref{fig:kdis}, and plotted dependent on the partial flow of nitrogen which is correlated to the partial pressure of nitrogen.

The error in $\rm k_{dis,N_2}$ estimation is also shown in figure \ref{fig:kdis} for each measurement point originating mostly from the error in pressure estimation which is higher for smaller nitrogen concentrations.

Besides one value for low nitrogen admixture, values for $k_\text{dis,N$_2$}$ range between $1.5-3.5\times 10^{-15}\,\mathrm{\frac{m^3}{s}} $, but drop a bit with increased flow of $\rm N_2$.
Nevertheless, values for higher nitrogen admixtures are considered trustworthier as they are less affected by errors of gas flows set at the mass flow controller (MFC), errors in estimation of total pressure and of 1s densities. Since the MFC works in the lower limit, the flow is not stable and has a higher variation. The comparison with the theoretical  estimation described in subsection \ref{disrate} for 340 K neutral gas temperature gives a rate constant of $\rm 2.59\times10^{-15}\, \frac{m^3}{s}$ and is in good agreement to our estimated values shown in figure \ref{fig:kdis}.

\section{Conclusion}

A method to evaluate TDLIF measurements accounting for intra- and inter-multiplet
transitions has been proposed for an argon/nitrogen magnetized plasma resulting in the evaluation of the disalignment rate coefficient for the $\rm 2p_8$ state of argon due to nitrogen molecular collisions.
Experimentally obtained results are found to be in good agreement with theoretical values.
The method was described for the $\rm 2p_8$ to $\rm 1s_4$, $\rm 1s_5$ and $\rm 1s_2$ schemes but can be easily
adopted for other 2p argon states. 
The proposed measurement method and analysis could further be used to evaluate other gas mixtures in magnetized plasma conditions not only limited to argon admixtures.
With the systematic evaluation of disalignment for argon 2p levels by other gas compounds and quantitative TDLIF, a basis for further development of an appropriate line branching method for evaluation of electron densities and temperature can be achieved for argon gas mixtures. These properties are making TDLIF measurements in magnetized plasma a highly efficient tool for quantitative plasma diagnostics while
providing basis for further understanding and description of light transport properties in argon gas mixtures.

\ack This work is supported by the Deutsche Forschungsgemeinschaft (DFG).
\section*{References}
\bibliographystyle{iopart-num}
\bibliography{paper}
\end{document}